# Pressure Induced Anomalous Metal in the Vicinity of the Superconductor Insulator Transition

Roy G. Cohen Song[1, *], Mark Nikolaevsky[1], Amitay Cohen[2], Ran Salem[2], Shamashis Sengupta[3], Claire A. Marrache-Kikuchi[3], and Aviad Frydman[1, *]
[1]*Department of Physics and Jack and Pearl Resnick Institute and Institute of Nanotechnology and Advanced Materials, Bar-Ilan University, Ramat-Gan 52900, Israel*
[2]*Physics Department, Nuclear Research Center - Negev, P.O. Box 9001, 84190 Be'er Sheva, Israel*
[3]*Université Paris-Saclay, CNRS/IN2P3, IJCLab, 91405 Orsay, France*

The Superconductor-to-Insulator Transition (SIT) in two-dimensional superconductors occurs due to a competition between superconductivity, quantum interferences, Coulomb interactions and disorder. Despite extensive theoretical and experimental investigation, the SIT remains an active research area due to the potential for exotic phases near the transition. One such phase is the Anomalous Metal, which has been claimed to exist between the insulating and superconducting states. This elusive phase, which is not consistent with current theories, is under heavy deliberations nowadays. We present an experimental study of the effect of high pressure on thin films of amorphous indium oxide. Our results show that pressure induces a series of transitions from a Bose insulator through a superconducting phase, metallic phases and finally to a conventional insulator. We suggest that our findings reaffirm the existence of a two-dimensional metal close to the SIT and show that its occurrence requires relatively strong coupling between regions that are weakly superconducting.

## I. INTRODUCTION

The interplay between superconductivity and disorder is a very active topic of investigation. It confronts two of the most important paradigms of condensed matter physics which are both very well understood and are macroscopic quantum phenomena: BCS superconductivity and Anderson localization [1]. And yet when these two come together, as realized in disordered 2D films, both paradigms break down across the phases leading to the "superconductor-insulator transition" (SIT). Experiments show that superconductivity in 2D films can exist but is destroyed by strong enough disorder [2–7] as well as by other non-thermal tuning parameters such as inverse thickness [8–19], magnetic field [20–30], chemical composition [31] and gate voltage [32, 33]. Once superconductivity is suppressed, it undergoes a transition to an insulating state (for reviews see [34, 35]). Since this transition occurs theoretically at $T = 0$, the SIT has been considered a very basic manifestation of a quantum phase transition, i.e. a phase transition which occurs at zero temperature as a function of some non-thermal parameter and is driven by quantum instead of thermal fluctuations. The importance of the SIT is both fundamental, since it offers insights into the relative strength of competing orders, and application-oriented, as low dimensional superconductors are the basis for many quantum engineering and detection applications (Qubits, bolometers, quantum sensors, etc).

Despite several decades of both theoretical and experimental extensive investigations, the SIT is more than ever an active field of research and discussions, among other reasons due to the possibility of exotic phases near the transition. One of these is the insulating phase which shows unconventional behavior. The electronic transport exhibits simple activation [36, 37] rather than the variable-range-hopping (VRH) behavior which is expected for disordered films. In addition, experiments have revealed signs for Cooper pairing [14, 18], vortex motion [4, 7] and finite superconducting energy gap [3, 5] in the insulating phase. Correspondingly, theories have predicted the presence of emergent electronic granularity [38–41] leading to superconducting islands embedded in an insulating matrix. Such a phase has been named the *Bosonic insulator* (BI)[42] and is associated with the presence of a pseudogap above the transition temperature or in the insulator.

Experiments and theories have also raised the possibility of another exotic phase, i.e. an intermediate anomalous Boson metal, between the insulator and the superconductor (for a review see [43]). The presence of a metallic phase in a two dimensional film contradicts the well-accepted scaling theory of localization [44]. Nevertheless, a wide variety of superconducting films, that are driven through the SIT using different tuning parameters, exhibit saturated resistance as the film is cooled to low temperatures in a large regime between the insulating and the superconducting phases, which has been considered an indication for metallic behavior [45–55].

This phase has been dubbed the *anomalous metal* (AM) not only because it is not consistent with the standard paradigms for transport in disordered 2D metals but also because the low-temperature conductivity can be orders of magnitude different than the expected Drude contribution $\sigma_D = ne^2\tau/m$. For films close to the superconducting state, as the film is cooled through a critical temperature, the resistance initially drops like a superconductor but eventually saturates at a value much smaller than $\rho_D$. For films close to the insulating phase, the resistivity initially increases like an insulator, but at low temperature it saturates at a value much larger than $\rho_D$ and can even be orders of magnitude larger than the quantum resistance of $h/e^2$. Following [51] we name the former behavior *metal phase 1* (*M1*) and the latter *metal phase 2* (*M2*).

* contact author: aviad.frydman@gmail.com

The AM state has been the subject of ongoing debate and it has generated a variety of explanations and models. Tamir et al. [56] showed that the metallic behavior that was seen in two different systems could be eliminated by better filtering external radiation. They claimed that superconductivity, at very low temperature in disordered films, is extremely sensitive to external noise, which suppresses the superconducting state, leading to resistance saturation, which can be interpreted as a metallic behavior. While this is definitely true in some cases, there is growing evidence that in many systems the AM is not just a result of poor noise filtering. Theories have suggested that the *M1* phase originates from quantum superconducting phase fluctuations (or vortices) which cause a finite voltage at very low temperatures [40, 55, 57–59], and the *M2* phase originates from quantum amplitude fluctuations (charge) leading to a finite current [55]. Alternatively, the AM was suggested to be the phase in which both charge and phase are disordered due to quantum fluctuations [60]. Another suggestion asserts that a novel vortex-glass state develops in disordered superconductors [61] where the system loses global phase coherence while retaining local phase coherence, thus becoming glassy as the phases of the order parameter are frozen along noncollinear directions leading to a metallic dissipative phase. Clearly, new experimental directions are required for shedding light on this elusive phase.

In this work, we introduce a novel tuning parameter for driving the SIT: high pressure. The effect of pressure $P$ on bulk superconductors has been extensively investigated [62]. Experiments have shown that pressure can either enhance superconductivity or destroy it. Many superconducting elements show a decrease of $T_c$ with pressure. This decrease is attributed to the volume dependence of the density of states at the Fermi energy, $N(E_F)$, and of the average phonon energy [63]. On the other hand, the $T_c$ of some high-critical temperature superconductors increases significantly with pressure [64]. The reasons suggested for this effect are diverse and heavily debated.

Here we study the effect of pressure on two-dimensional superconductivity. By applying pressure in the 10 GPa range to thin layers of amorphous indium oxide (InO) we are able to tune the films through the different phases of the SIT. Our main results are as follows:

1. Pressure can drive a system through the SIT but in an unusual way. A sample which begins as a BI transits to a superconductor and then back to an insulator as a function of increasing pressure.

2. The high pressure insulator is different in nature than the ambient pressure one. While the latter exhibits simple activation transport typical of a BI, the former shows VRH behavior characteristic of a conventional Anderson insulator.

3. Intermediate anomalous metals of both *M1* and *M2* types are observed between the superconducting and insulating states in InO. This is particularly intriguing since, in this material, when the SIT is driven by other tuning parameters, metallic behavior is not observed.

Our results provide valuable information about the physics of disordered superconductors in the 2D limit and especially assist in illuminating the conditions for observing the AM phase in these systems.

## II. RESULTS

For this study, we employed a unique technique which we developed for pressurizing ultra-thin films [65], as described in the Methods section (see Fig. 1).

We measured seven thin InO films which were prepared under different partial $0_2$ pressure during deposition such that the samples at $P = 0$ span the SIT. The film thickness, 30 nm, was comparable to the superconducting coherence length, $\xi$ [66], placing all seven samples in the 2D superconducting limit. Resistance versus temperature measurements (RT) were performed incrementally as $P$ was increased and reached a maximum pressure as high as 17 GPa (before contacts were lost or the diamond broke). Fig. 2 shows examples for RTs at different pressures for three samples. *Sample I* is an insulator (a), *Sample T* is close to the SIT (b) and *Sample S* is a superconductor (c) at $P = 0$ GPa. Similar results (see supplementary material S1) were obtained for all studied samples which were measured either in a $^3$He cryostat or in a dilution refrigerator, both of which were adequately filtered from external radiation.

For *Sample I*, pressure first induces a transition from an insulator to a superconductor, then to an AM, and finally to a weakly insulating state (Fig. 2a). Similarly, for *sample T*, an initial pressure increase (up to ∼

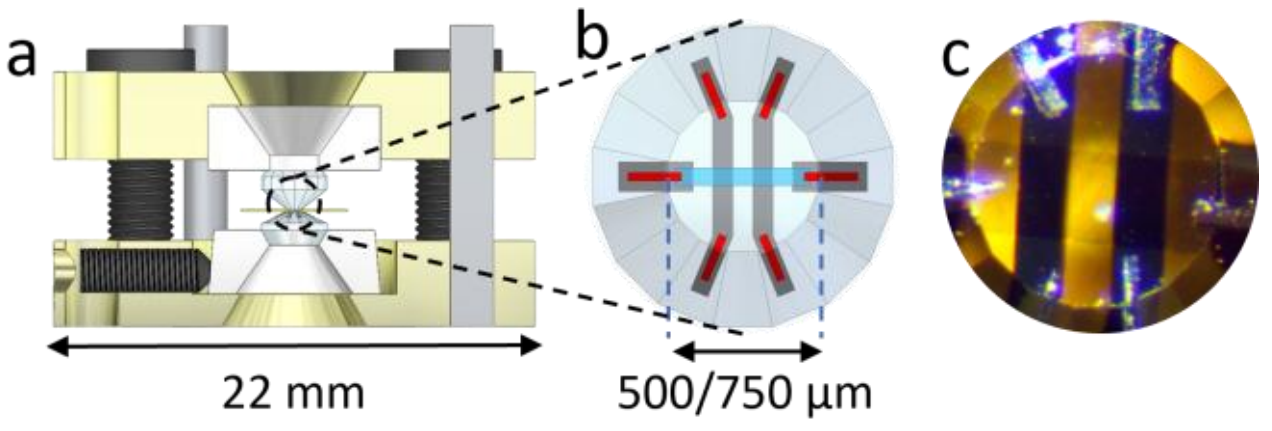

FIG. 1. **The experimental setup. (a)** Schematic description of our opposing plate DAC. **(b)** Sketch of the sample deposited on the bottom diamond. The gray strips are the evaporated Pt leads, the cyan strip is the sample and the red patches are the Pt foils. **(c)** A microscope image of a fabricated sample.

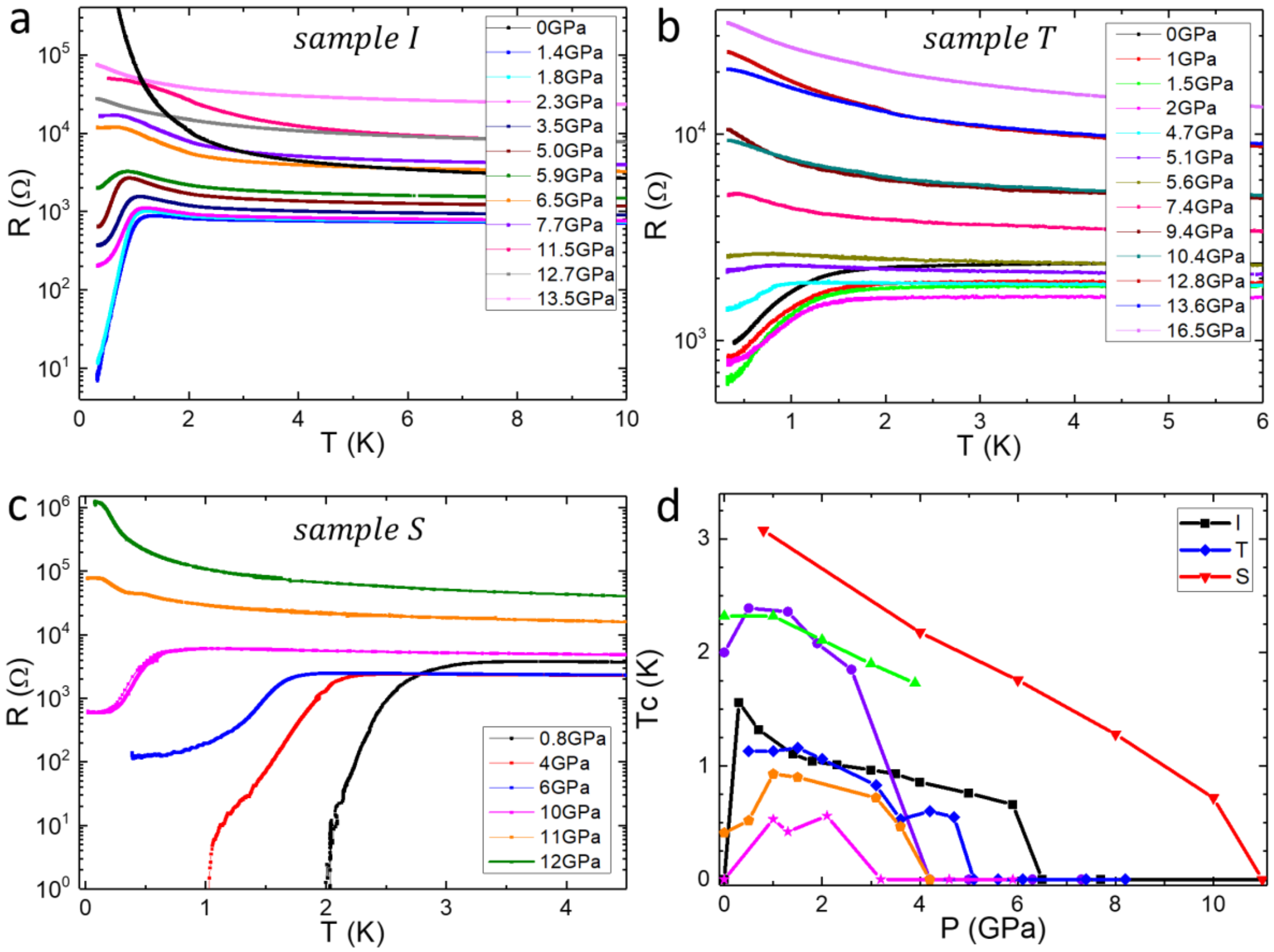

FIG. 2. **Resistance versus temperature at different pressures for three representative samples: (a)** a sample which is insulating at ambient pressure (Sample I), **(b)** a sample which is close to the transition at ambient pressure (Sample T) and **(c)** a sample which is superconducting at ambient pressure (Sample S). The three samples show a metallic state in between the superconducting and the insulating states of the SIT. **(d)** Superconducting critical temperature, Tc, defined as the temperature at which the resistance drops to 90% of its maximal value, as a function of pressure, for all measured samples.

3 GPa) strengthens superconductivity while still in the *M1* phase, and higher pressure suppresses superconductivity until, eventually the film becomes insulating (Fig. 2a-c). This trend can be seen in Fig. 2d which depicts $T_c$ as a function of pressure for the seven samples. Here, $T_c$ is defined as the temperature at which the resistance dropped to 90% of its maximal value, thus being relevant both for the S and for the *M2* phases. A non-monotonic behavior of $T_c$ is observed for nearly all samples. Notably, samples which are prepared initially deeper in the superconducting state show a monotonic decrease of $T_c$ before the sample switches to an insulating state as is the case for *Sample S* (Fig. 2c.)

Perhaps the most striking finding is the observation of metallic behaviors of both *M1* and *M2* nature in all of our films between the S and the I phases. These are demonstrated by the low temperature resistance saturation at intermediate $P$ as depicted in Fig. 2a-c. It is important to note that we have not observed such metallic behaviors in InO samples measured at ambient pressure, both in this work and in the past, using the same experimental setups. Furthermore, resistance saturation is observed at very different temperatures, as high as 1 K, so that this phenomenon is unlikely to arise from heating or noise issues. Most importantly, the same metallic behavior was observed using two different setups in two different labs on the same sample. All these facts point to a genuine pressure-induced metallic ground state in these InO films.

## III. DISCUSSION

In a search for the physical origin of the intermediate metallic phase and the role played by pressure in generating it, we note that the non-monotonic trend of superconductivity versus pressure shown in Fig. 2d points to a competition between two phenomena, one which tends to increase $T_c$, the other which suppresses it.

To understand this, we first turn our attention to the behavior of *sample I* which is initially (at $P = 0$) insulating (Fig. 2a). This insulator, dubbed *I1* shows simple activation transport: $R \propto exp[1/T]$ as demonstrated in Fig. 3a. Such behavior, which is typical of ambient pressure insulating InO [22, 36], has been attributed to emergent granularity in thin samples close to the SIT [38, 41]. In this picture, disorder on a microscopic length-scale can induce superconducting granularity on a mesoscopic scale when the sample is cooled below $T_c$. Local superconductivity within the grains determines the energy gap, Δ, while global superconductivity, superfluid density and $T_c$ are governed by the coupling between the grains [67]. The activated behavior of the RT in the insulating phase can thus be attributed to electrons hopping between grains and having to overcome both the charging energy and the superconducting gap [37].

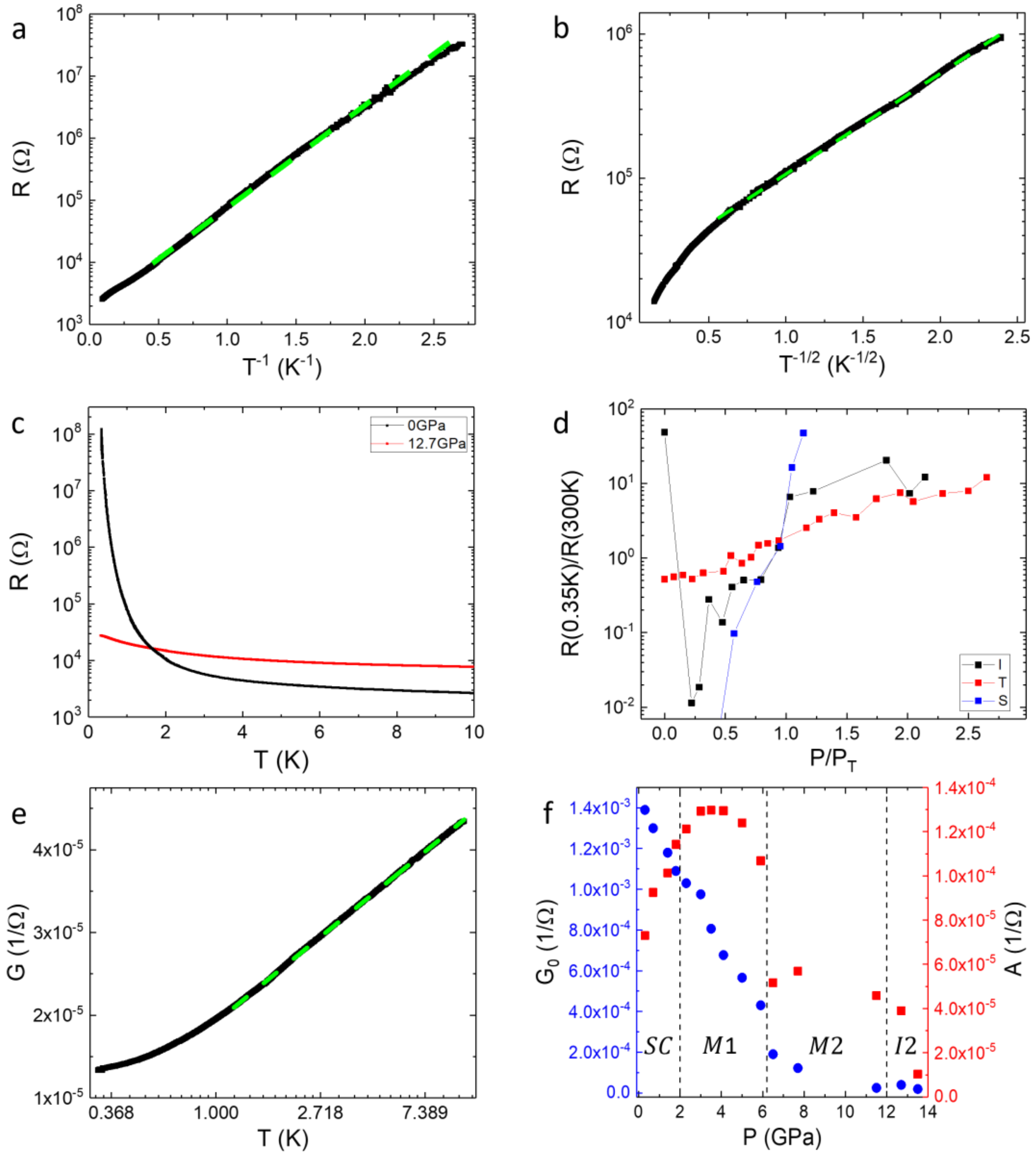

FIG. 3. **From a Bose insulator to a Fermi insulator (a)** Resistance versus 1/T of Sample I at ambient pressure showing simple activation behavior typical of a BI. **(b)** Resistance versus T −1/2 of Sample S at P = 12 GPa. The straight line corresponds to Efros-Shklovskii type VRH, typical of an Anderson insulator. **(c)** Resistance versus temperature of Sample I at ambient pressure and 12.7 GPa showing clear crossing of the curves. **(d)** Ratio of the resistance at 350 mK and at 300 K for the three samples of Fig. 2 versus the pressure normalized to the critical pressure PT where the sample transits from M1 to M2. **(e)** Typical conductance versus log(T ) of the 13.5 GPa stage of Sample I exhibiting weak localization between 1 and 10 K. The dashed line is a fit to Eq. 1. **(f)** $G_0$ and A as a function of pressure extracted from such fits for Sample I. Similar results for other samples are shown in supplementary material S2 and S3.

Under application of relatively low pressure, the initially insulating film becomes superconducting. This is a clear indication that pressure, at least initially, increases the coupling between superconducting grains. A similar conclusion can be drawn from the fact that application of pressure below 1.5 GPa on *Sample T* causes an increase of $T_c$. On the other hand, the fact that, for all samples, a high enough pressure causes a decrease of $T_c$ as clearly seen in Fig. 2d, shows that, eventually, pressure acts to reduce superconductivity.

These findings suggests that the pressure plays a double role on disordered superconductors which contain emergent superconducting regions. On one hand pressure decreases $T_c$ in each island, similar to the effect of pressure on conventional superconducting elements [63]. On the other hand, pressure enhances the coupling between different superconducting regions thus increasing Josephson coupling and global phase coherence. This leads to the non-monotonic trend of $T_c$.

This scenario is supported by two additional findings. First, at high pressure, all samples transits to an insulating phase, *I2*, which, unlike *I1*, is not characterized by activated transport but rather shows conventional variable range hopping $R \propto exp[1/T]^{\alpha}$ with $\alpha = 1/2$ (Fig. 3b). This implies that *I2* is a usual fermionic Anderson insulator indicating that the nature of the insulator changes with increasing pressure. Indeed, the fact that RTs cross at intermediate temperature is highly unusual in these films. It means that a higher disorder (i.e. higher room temperature resistance) gives rise to a weaker insulator at low temperature. This kind of crossing can be seen for the RTs of *I1* and *I2* (Fig. 2c), but are also observed in *Sample T* between the curves at 9.4 and 10.4 GPa and between 12.8 and 13.6 GPa. In the latter case, the system's disorder only increases very slightly, but the residual resistance decreases. A similar conclusion can be drawn from the non-monotonic dependence of the inverse of the residual resistive ratio (RRR) on the pressure (Fig. 3d.)

The second finding supporting our scenario emerges from an analysis of the transport properties above $T_c$. We fit the conductance versus temperature curves above $T_c$ to the 2D quantum corrections to Drude conductivity (due to weak localization (WL) and electron-electron interactions) using the following expression:

$$G = G_0 + A \times \ln\left(\frac{T}{T_0}\right) \quad (1)$$

(see Fig.3e for one stage of *Sample I*). Here $G_0$ is the Drude conductivity which decreases with disorder, $T_0$ is the temperature at which the conductivity diverts from the Drude expectation which is expected to decrease with disorder and $A$ has been shown experimentally to increase as the sample becomes more localized. Fig 3f shows that $G_0$ decreases monotonically, indicating increasing disorder, with increasing pressure. However, for a large regime of increasingly high pressure, $A$ is also suppressed, thus suggesting decrease of localization effects. This weakening of the temperature dependence of conductivity despite the increasing value of resistance, which is seen in all samples, is very counterintuitive and supports the understanding that the nature of the insulator is being modified with temperature. Similar behavior is found for all samples (see supplementary material).

We also note that, for all samples, the transition from *M2* to *I2* occurs at A of the order of $1\times10^{-5}$ S, which is close to the theoretical value given by the weak localization theory: $A = \frac{p}{2}\frac{e^2}{\pi^2\hbar}$, with p = 3 if localization is due to electron-phonon interactions, p = 3/2 or 1 for electron-electron interactions in 3D or 2D respectively [68].

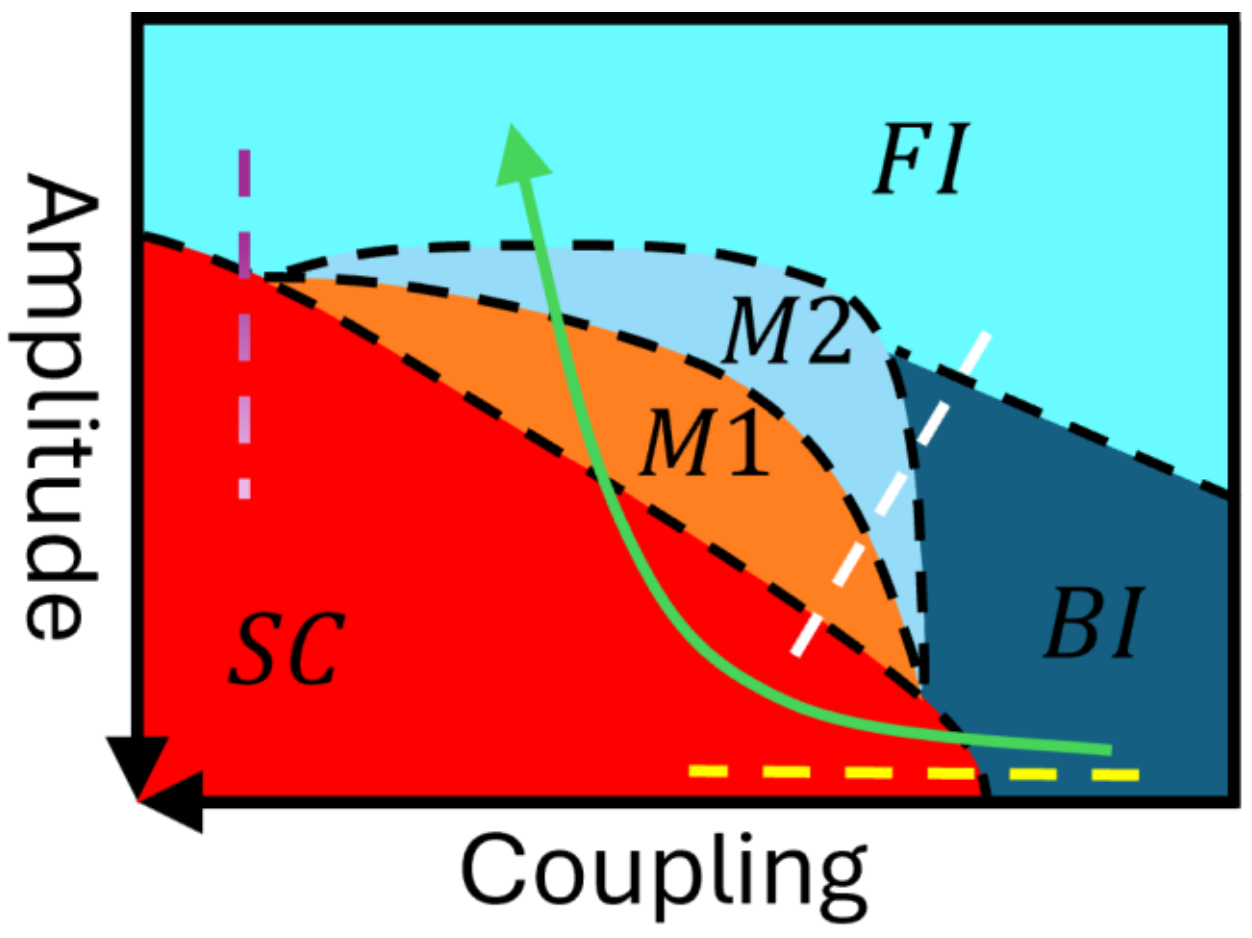

FIG. 4. **Schematic of the envisioned phase diagram**. The diagram shows the different phases: superconductor, *I1* (Bose insulator), *I2* (Fermionic insulator), *M1* and *M2*. The two axes are the superconducting order parameter amplitude and phase coupling. The green arrow shows the effect of increasing pressure. The dashed lines are example of different kinds of SIT at ambient pressure, yellow for a direct superconductor to Bose insulator as seen in ambient InO [7], purple for a direct superconductor to Fermion insulator as perhaps seen in ultra-thin metals such as Bi [10], and white for a transition that includes metallic states as seen e.g. in NbSi [51].

As previously noted, all samples exhibit the *M1* and *M2* states between the *S* and *I2* phases. It is worth recalling that ambient pressure InO samples never display AM features; these states emerge only under applied pressure. This suggests that the formation of metallic phases requires very specific conditions. In particular, the AM seems to appear when two conditions are met: superconductivity within each grain is significantly suppressed by pressure, and the grains are strongly coupled to one another. Based on this, we propose a schematic phase diagram in Fig. 4, to illustrate the role of both the amplitude and phase of the superconducting order parameter in generating the various phases of disordered 2D films. The appearance of the AM phase depends critically on strong coupling between regions where the superconducting order parameter amplitude is highly suppressed. This is consistent with the fact that most sys-tems that have been reported to systematically exhibit AM properties, such as NbSi, are characterized by a relatively low maximal $T_c$ indicating low superconducting order parameter amplitude.

## IV. CONCLUSION

By applying pressure to InO films, we were able to induce metallic behaviors in a material that does not exhibit such behavior otherwise. The persistence of resistance saturation up to a temperature of $\approx 1$ K, similarly in two different setups, further reinforces the notion that this behavior is not due to heating or noise. Although the origin of the anomalous metallic phase in various 2D superconductors remains poorly understood, our findings provide crucial insights into the conditions necessary for its emergence. We hope these results will stimulate further theoretical efforts aimed at fully characterizing this elusive state of matter.

## METHODS

Our measurement setup is based on an opposing plate diamond anvil cell [69] (DAC) made of CuBe (Copper Beryllium) to ensure good thermal conductivity between the thermal bath and the sample. This material is also strong enough to withstand $10-20$ GPa. The DAC for this work was equipped with Brilliant cut and Bohler-Almax (BA) design anvils with a culet size of 500/750 $\mu$m. A pre-indented CuBe gasket with 300-500 $\mu$m hole was pressed between the diamonds. We covered the gasket with a mixture of epoxy glue and MgO to prevent electrical shorts between the leads. The same epoxy (without MgO) was also used as a pressure medium.

Thin films were grown directly on one diamond. Using a mechanical mask, platinum contacts were sputtered to form electrical leads going from the culet to the bottom of the bevel (Fig 1b). An amorphous indium oxide film was then deposited using e-gun evaporation of InO through a rectangular mask placed above the center of the culet between the contacts. High-purity oxygen gas was inserted into the chamber at a partial pressure of $1-5\times10^{-5}$ mbar. This partial pressure determined the initial ground state of the film (insulator of superconductor) [7, 70]. Since thermal anchoring the diamond to the DAC is difficult to achieve at low temperatures and ambient pressure, we simultaneously evaporated a test sample on a free diamond for ambient pressure measurements. To ensure electrical conductivity at high pressure, 7 $\mu$m-thick Platinum foils were added on top of the evaporated platinum contacts after the film deposition and held in place by silver epoxy. Pressure was determined by measuring the spectroscopy of Ruby spheres that were placed between the two diamonds [71]. All measurements were preformed either in a $^3$He cryostat or in a dilution fridge using standard low-noise measurement techniques.

## V. ACKNOWLEDGMENTS

We gratefully acknowledge the Paris-Saclay University Jean D'Alembert fellowship (AF). This project was partially funded by the the Pazy foundation and partially by the CNRS International Emerging Action (grant IEA C100195) and by a France 2030 funding (PhOM - Graduate School Physique, grant ANR-11-IDEX-0003).

## AUTHORS CONTRIBUTION

RGCS prepared the samples. RGCS, SS and CMK performed the experiments. MN, AC and RS provided knowledge and supplies for high pressure experiments, AF initiated the project and supervised it together with CMK. RGCS, CMK and AF wrote the manuscript.

## COMPETING INTERESTS

There are no competing interests to declare.

## DATA AND MATERIALS AVAILABILITY

All data are available upon reasonable request at royg.cohen@gmail.com.

[1] P.W. Anderson. Absence of diffusion in certain random lattices. *Physical review*, 109(5):1492, 1958.

[2] B Sacépé, C Chapelier, T.I. Baturina, V.M. Vinokur, M.R. Baklanov, and M Sanquer. Disorder-induced inhomogeneities of the superconducting state close to the superconductor-insulator transition. *Physical Review Letters*, 101(15):157006, 2008.

[3] B Sacépé, T Dubouchet, C Chapelier, M Sanquer, M Ovadia, D Shahar, M Feigel'Man, and L Ioffe. Localization of preformed cooper pairs in disordered superconductors. *Nature Physics*, 7(3):239–244, 2011.

[4] S Poran, E Shimshoni, and A Frydman. Disorder-induced superconducting ratchet effect in nanowires. *Physical Review B—Condensed Matter and Materials Physics*, 84(1):014529, 2011.

[5] D Sherman, G Kopnov, D Shahar, and A Frydman. Measurement of a superconducting energy gap in a homogeneously amorphous insulator. *Physical Review Letters*, 108(17):177006, 2012.

[6] O Crauste, A Gentils, F Couëdo, Y Dolgorouky, L Bergé, S Collin, C.A. Marrache-Kikuchi, and L Dumoulin. Effect of annealing on the superconducting properties of a-nb x

si 1-x thin films. *Physical Review B—Condensed Matter and Materials Physics*, 87(14):144514, 2013.

[7] A Roy, E Shimshoni, and A Frydman. Quantum criticality at the superconductor-insulator transition probed by the nernst effect. *Physical Review Letters*, 121(4):047003, 2018.

[8] M Strongin, R.S. Thompson, O.F. Kammerer, and J.E. Crow. Destruction of superconductivity in disordered near-monolayer films. *Physical Review B*, 1(3):1078, 1970.

[9] R.C. Dynes, A.E. White, J.M. Graybeal, and J.P. Garno. Breakdown of eliashberg theory for two-dimensional superconductivity in the presence of disorder. *Physical Review Letters*, 57(17):2195, 1986.

[10] D.B. Haviland, Y Liu, and A.M. Goldman. Onset of superconductivity in the two-dimensional limit. *Physical Review Letters*, 62(18):2180, 1989.

[11] R.C. Valles Jr, J.M.and Dynes and J.P. Garno. Electron tunneling determination of the order-parameter amplitude at the superconductor-insulator transition in 2d. *Physical Review Letters*, 69(24):3567, 1992.

[12] A Frydman, O Naaman, and R.C. Dynes. Universal transport in two-dimensional granular superconductors. *Physical Review B*, 66(5):052509, 2002.

[13] A Frydman. The superconductor insulator transition in systems of ultrasmall grains. *Physica C: Superconductivity*, 391(2):189–195, 2003.

[14] M.D. Stewart Jr, A Yin, J.M. Xu, and J.M. Valles Jr. Superconducting pair correlations in an amorphous insulating nanohoneycomb film. *Science*, 318(5854):1273–1275, 2007.

[15] C.A. Marrache-Kikuchi, H Aubin, A Pourret, K Behnia, J Lesueur, L Bergé, and L Dumoulin. Thickness-tuned superconductor-insulator transitions under magnetic field in a-nbsi. *Physical Review B—Condensed Matter and Materials Physics*, 78(14):144520, 2008.

[16] S.M. Hollen, H.Q. Nguyen, E Rudisaile, M.D. Stewart Jr, J Shainline, J.M. Xu, and J.M. Valles Jr. Cooper-pair insulator phase in superconducting amorphous bi films induced by nanometer-scale thickness variations. *Physical Review B—Condensed Matter and Materials Physics*, 84(6):064528, 2011.

[17] T.I. Baturina, V.M. Vinokur, A.Y. Mironov, N.M. Chtchelkatchev, D.A. Nasimov, and A.V. Latyshev. Nanopattern-stimulated superconductor-insulator transition in thin tin films. *Europhysics Letters*, 93(4):47002, 2011.

[18] G Kopnov, O Cohen, M Ovadia, K.H. Lee, C.C. Wong, and D Shahar. Little-parks oscillations in an insulator. *Physical Review Letters*, 109(16):167002, 2012.

[19] S Poran, Ml Molina-Ruiz, A Gérardin, A Frydman, and O Bourgeois. Specific heat measurement set-up for quench condensed thin superconducting films. *Review of Scientific Instruments*, 85(5), 2014.

[20] M.A. Paalanen, A.F. Hebard, and R.R. Ruel. Low-temperature insulating phases of uniformly disordered two-dimensional superconductors. *Physical Review Letters*, 69(10):1604, 1992.

[21] A Yazdani and A Kapitulnik. Superconducting-insulating transition in two-dimensional a-moge thin films. *Physical Review Letters*, 74(15):3037, 1995.

[22] G Sambandamurthy, L.W. Engel, A Johansson, and D Shahar. Superconductivity-related insulating behavior. *Physical Review Letters*, 92(10):107005, 2004.

[23] G Sambandamurthy, L.W. Engel, A Johansson, E Peled, and D Shahar. Experimental evidence for a collective insulating state in two-dimensional superconductors. *Physical Review Letters*, 94(1):017003, 2005.

[24] M.A. Steiner, G Boebinger, and A Kapitulnik. Possible field-tuned superconductor-insulator transition in high-t c superconductors: Implications for pairing at high magnetic fields. *Physical Review Letters*, 94(10):107008, 2005.

[25] T.I. Baturina, J Bentner, C Strunk, M.R. Baklanov, and A Satta. From quantum corrections to magnetic-field-tuned superconductor–insulator quantum phase transition in tin films. *Physica B: Condensed Matter*, 359:500–502, 2005.

[26] H Aubin, C.A. Marrache-Kikuchi, A Pourret, K Behnia, L Berg´e, L Dumoulin, and J Lesueur. Magnetic-field-induced quantum superconductor-insulator transition in nb 0.15 si 0.85. *Physical Review B—Condensed Matter and Materials Physics*, 73(9):094521, 2006.

[27] T.I. Baturina, C Strunk, M.R. Baklanov, and A Satta. Quantum metallicity on the high-field side of the superconductor-insulator transition. *Physical Review Letters*, 98(12):127003, 2007.

[28] R.W. Crane, N.P. Armitage, A Johansson, G Sambandamurthy, D Shahar, and G Grüner. Fluctuations, dissipation, and nonuniversal superfluid jumps in two-dimensional superconductors. *Physical Review B—Condensed Matter and Materials Physics*, 75(9):094506, 2007.

[29] V M Vinokur, T.I. Baturina, M.V. Fistul, A.Y. Mironov, M.R. Baklanov, and C Strunk. Superinsulator and quantum synchronization. *Nature*, 452(7187):613–615, 2008.

[30] R Ganguly, I Roy, A Banerjee, H Singh, A Ghosal, and P Raychaudhuri. Magnetic field induced emergent inhomogeneity in a superconducting film with weak and homogeneous disorder. *Physical Review B*, 96(5):054509, 2017.

[31] M Mondal, A Kamlapure, M Chand, G Saraswat, S Kumar, J Jesudasan, L Benfatto, V Tripathi, and P Raychaudhuri. Phase fluctuations in a strongly disordered s-wave nbn superconductor close to the metal-insulator transition. *Physical Review Letters*, 106(4):047001, 2011.

[32] K.A Parendo, K.S.B Tan, A Bhattacharya, M Eblen-Zayas, N.E. Staley, and A.M. Goldman. Electrostatic tuning of the superconductor-insulator transition in two dimensions. *Physical Review Letters*, 94(19):197004, 2005.

[33] AD Caviglia, S Gariglio, N Reyren, D Jaccard, T Schneider, M Gabay, S Thiel, G Hammerl, J Mannhart, and J.M. Triscone. Electric field control of the laalo3/srtio3 interface ground state. *Nature*, 456(7222):624–627, 2008.

[34] A.M. Goldman and N Markovi´c. Superconductor-insulator transitions in the two-dimensional limit, 1998.

[35] Y.H. Lin, J Nelson, and A.M. Goldman. Superconductivity of very thin films: The superconductor–insulator transition. *Physica C: Superconductivity and its Applications*, 514:130–141, 2015.

[36] D Shahar and Z Ovadyahu. Superconductivity near the mobility edge. *Physical Review B*, 46(17):10917, 1992.

[37] V Humbert, M Ortuño, A.M. Somoza, L Berg´e, L Dumoulin, and C.A. Marrache-Kikuchi. Overactivated transport in the localized phase of the superconductor-insulator transition. *Nature communications*, 12(1):6733, 2021.

[38] A Ghosal, M Randeria, and N Trivedi. Role of spatial amplitude fluctuations in highly disordered s-wave superconductors. *Physical Review Letters*, 81(18):3940, 1998.

[39] A Ghosal, M Randeria, and N Trivedi. Inhomogeneous pairing in highly disordered s-wave superconductors. *Physical Review B*, 65(1):014501, 2001.

[40] E Shimshoni, A Auerbach, and A Kapitulnik. Transport through quantum melts. *Physical Review Letters*, 80(15):3352, 1998.

[41] K Bouadim, Y.L. Loh, M Randeria, and N Trivedi. Single-and two-particle energy gaps across the disorder-driven superconductor–insulator transition. *Nature Physics*, 7(11):884–889, 2011.

[42] M.P. Fisher and D.H. Lee. Correspondence between two-dimensional bosons and a bulk superconductor in a magnetic field. *Physical Review B*, 39(4):2756, 1989.

[43] A Kapitulnik, S.A. Kivelson, and B Spivak. Colloquium: anomalous metals: failed superconductors. *Reviews of Modern Physics*, 91(1):011002, 2019.

[44] E Abrahams, P.W. Anderson, D.C. Licciardello, and T.V. Ramakrishnan. Scaling theory of localization: Absence of quantum diffusion in two dimensions. *Physical Review Letters*, 42(10):673, 1979.

[45] A.F. Hebard and M.A. Paalanen. Magnetic-field-tuned superconductor-insulator transition in two-dimensional films. *Physical Review Letters*, 65(7):927, 1990.

[46] H.M. Jaeger, D.B. Haviland, B.G. Orr, and A.M. Goldman. Onset of superconductivity in ultrathin granular metal films. *Physical Review B*, 40(1):182, 1989.

[47] D Ephron, A Yazdani, A Kapitulnik, and M.R. Beasley. Observation of quantum dissipation in the vortex state of a highly disordered superconducting thin film. *Physical Review Letters*, 76(9):1529, 1996.

[48] C.D. Chen, P Delsing, D.B. Haviland, Y Harada, and T Claeson. Flux flow and vortex tunneling in two-dimensional arrays of small josephson junctions. *Physical Review B*, 54(13):9449, 1996.

[49] N Mason and A Kapitulnik. Dissipation effects on the superconductor-insulator transition in 2d superconductors. *Physical Review Letters*, 82(26):5341, 1999.

[50] C Christiansen, L.M. Hernandez, and A.M. Goldman. Evidence of collective charge behavior in the insulating state of ultrathin films of superconducting metals. *Physical Review Letters*, 88(3):037004, 2002.

[51] F Cou¨edo, Or Crauste, A.A. Drillien, V Humbert, L Berg´e, C.A. Marrache-Kikuchi, and L Dumoulin. Dissipative phases across the superconductor-to-insulator transition. *Scientific Reports*, 6(1):35834, 2016.

[52] C.G.L. Bøttcher, F Nichele, M Kjaergaard, H.J. Suominen, J Shabani, C.J. Palmstrøm, and C.M. Mar- cus. Superconducting, insulating and anomalous metal- lic regimes in a gated two-dimensional semiconductor–superconductor array. *Nature Physics*, 14(11):1138–1144, 2018.

[53] Z Chen, A G Swartz, H Yoon, H Inoue, T A Merz, D Lu, Y Xie, H Yuan, Y Hikita, S Raghu, et al. Carrier density and disorder tuned superconductor-metal transition in a two-dimensional electron system. *Nature communications*, 9(1):4008, 2018.

[54] Chao Yang, Yi Liu, Yang Wang, Liu Feng, Qianmei He, Jian Sun, Yue Tang, Chunchun Wu, Jie Xiong, Wanli Zhang, et al. Intermediate bosonic metallic state in the superconductor-insulator transition. *Science*, 366(6472):1505–1509, 2019.

[55] X Zhang, A Palevski, and A Kapitulnik. Anomalous metals: From “failed superconductor” to “failed insulator”. *Proceedings of the National Academy of Sciences*, 119(29):e2202496119, 2022.

[56] I Tamir, A Benyamini, E.J. Telford, F Gorniaczyk, A Doron, T Levinson, D Wang, F Gay, B Sac´ep´e, J Hone, et al. Sensitivity of the superconducting state in thin films. *Science advances*, 5(3):eaau3826, 2019.

[57] B Spivak, Ar Zyuzin, and M Hruska. Quantum superconductor-metal transition. *Physical Review B*, 64(13):132502, 2001.

[58] B Spivak, P Oreto, and S.A. Kivelson. Theory of quantum metal to superconductor transitions in highly conducting systems. *Physical Review B—Condensed Matter and Materials Physics*, 77(21):214523, 2008.

[59] K Ienaga, T Hayashi, Y Tamoto, S Kaneko, and S Okuma. Quantum criticality inside the anomalous metallic state of a disordered superconducting thin film. *Physical Review Letters*, 125(25):257001, 2020.

[60] D Das and S Doniach. Existence of a bose metal at t= 0. *Physical Review B*, 60(2):1261, 1999.

[61] P Phillips and D Dalidovich. The elusive bose metal. *Science*, 302(5643):243–247, 2003.

[62] B Lorenz and C.W. Chu. High pressure effects on superconductivity. In *Frontiers in Superconducting Materials*, pages 459–497. Springer, 2005.

[63] R.E. Hodder. Pressure effects on the superconducting transition temperature of pb. *Physical Review*, 180(2):530, 1969.

[64] L Gao, Y.Y. Xue, F Chen, Q Xiong, R.L. Meng, D Ramirez, C.W. Chu, J.H. Eggert, and H.K. Mao. Superconductivity up to 164 k in $hgba_2ca_{m-1}cu_mo_{2m+2+\delta}$ (m= 1, 2, and 3) under quasihydrostatic pressures. *Physical Review B*, 50(6):4260, 1994.

[65] R Cohen, M Nikolaevsky, R Salem, and A Fryd- man. Setup for pressurizing thin films through the superconductor–insulator transition. *Review of Scientific Instruments*, 92(8), 2021.

[66] A Johansson, G Sambandamurthy, D Shahar, N Jacobson, and R Tenne. Nanowire acting as a superconducting quantum interference device. *Physical Review Letters*, 95(11):116805, 2005.

[67] D Sherman, B Gorshunov, S Poran, N Trivedi, E Farber, M Dressel, and A Frydman. Effect of coulomb interactions on the disorder-driven superconductor-insulator transition. *Physical Review B*, 89(3):035149, 2014.

[68] Patrick A Lee and Tiruppattur V Ramakrishnan. Disordered electronic systems. *Reviews of Modern Physics*, 57(2):287, 1985.

[69] A Jayaraman. Diamond anvil cell and high-pressure physical investigations. *Reviews of Modern Physics*, 55(1):65, 1983.

[70] Z Ovadyahu. Some finite temperature aspects of the anderson transition. *Journal of Physics C: Solid State Physics*, 19(26):5187, 1986.

[71] H.K. Mao, J.A. Xu, and P.M. Bell. Calibration of the ruby pressure gauge to 800 kbar under quasi-hydrostatic conditions. *Journal of Geophysical Research: Solid Earth*, 91(B5):4673–4676, 1986.

## SUPPLEMENTARY MATERIAL

# Pressure induced Anomalous Metal in the vicinity of the Superconductor Insulator Transition - Supplementary information

## I. TRANSPORT MEASUREMENT OF OTHER SAMPLES

We have measured seven InO films under various pressures. All samples showed the same behavior: the initial pressure increase strengthened the superconductivity, but higher pressure destroyed it. The samples undergo a transition from *I1*, a simply activated insulator, to superconductivity, *M1*, *M2* and, in the end, to *I2*, a Variable Range Hopping-type insulator. Figure S1 shows the resistance versus temperature curves for all samples that were not shown in the main text.

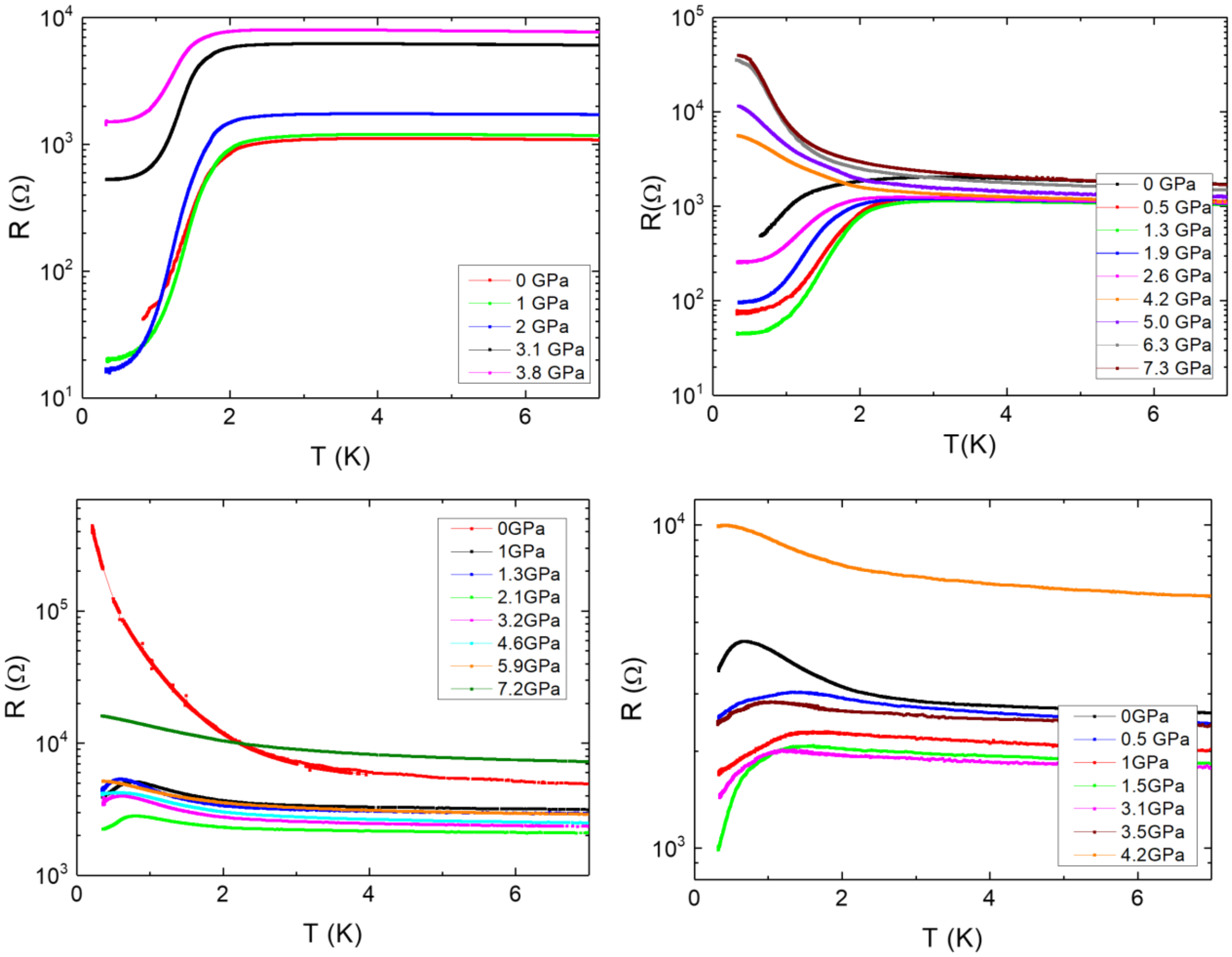

FIG. S1: Resistance versus temperature at different pressures for four InO films.

## II. WEAK LOCALIZATION

We fit the conductance versus temperature curves above typically 1 K to the 2D quantum corrections to Drude conductivity (due to weak localization (WL) and electron-electron interactions) using the following expression:

$$G = G_0 + A \times ln(T/T_0). \quad \text{(S1)}$$

$G_0$, is the Drude conductivity which decreases with disorder, $T_0$ is the temperature at which the conductivity diverts from the Drude expectation which is expected to decrease with disorder and $A$ has been shown experimentally to increase as the sample becomes more localized. Figure S2 shows the fitting to equation S1 for *Sample I* at different pressure stages. The corresponding fitting parameters $A$ and $G_0$ are given for *Sample T* and *Sample S* in figure S3 (they were given in figure 3.f of the main text for *Sample I*). As pressure increases, $G_0$ decreases monotonically, indicating increasing disorder. Eventually, at high pressure, $A$ is suppressed, thus suggesting weaker localization.

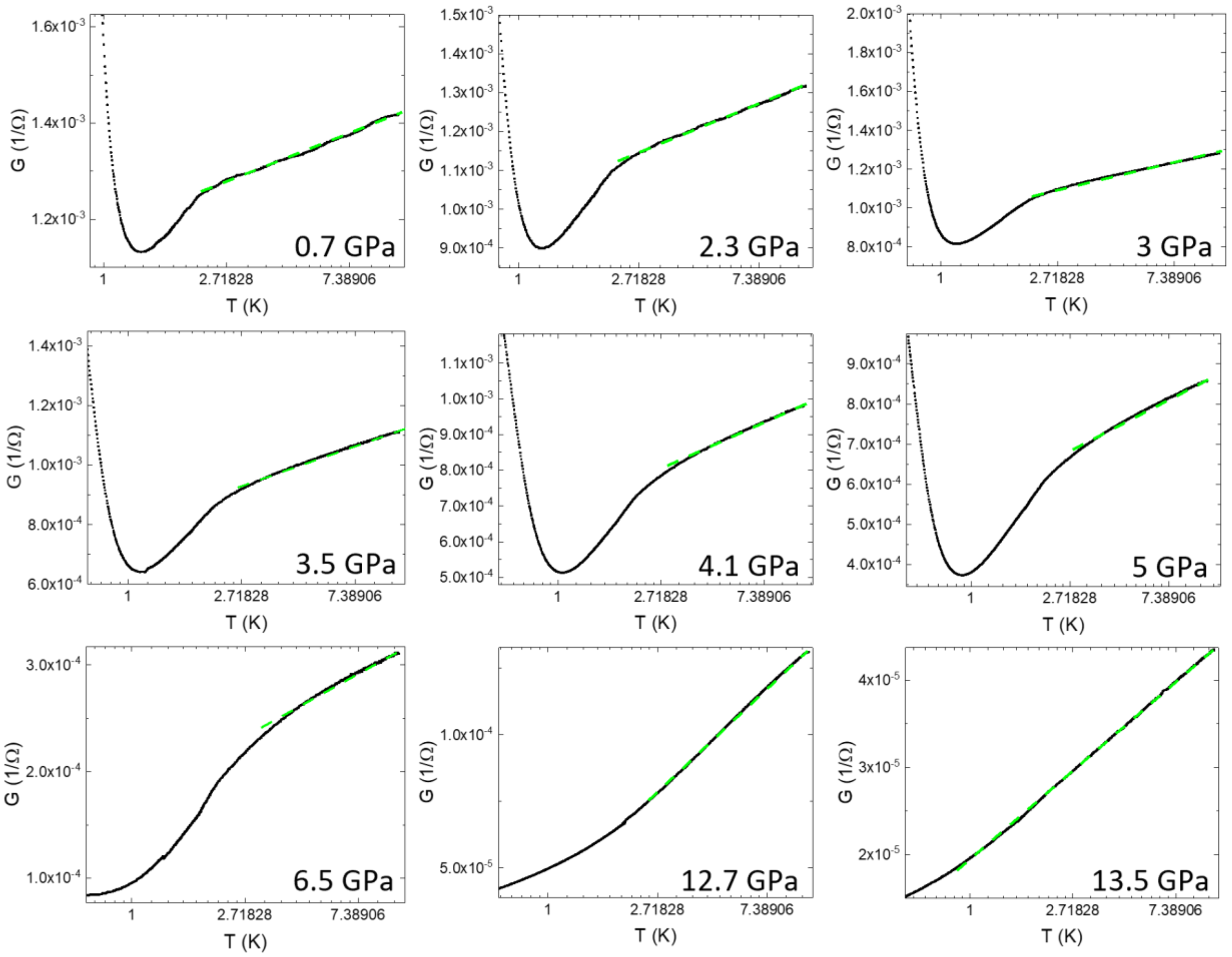

FIG. S2: Conductance versus log($T$) for various pressure stages of *Sample I* exhibiting weak localization. The dashed line is a fit to Eq. S1.

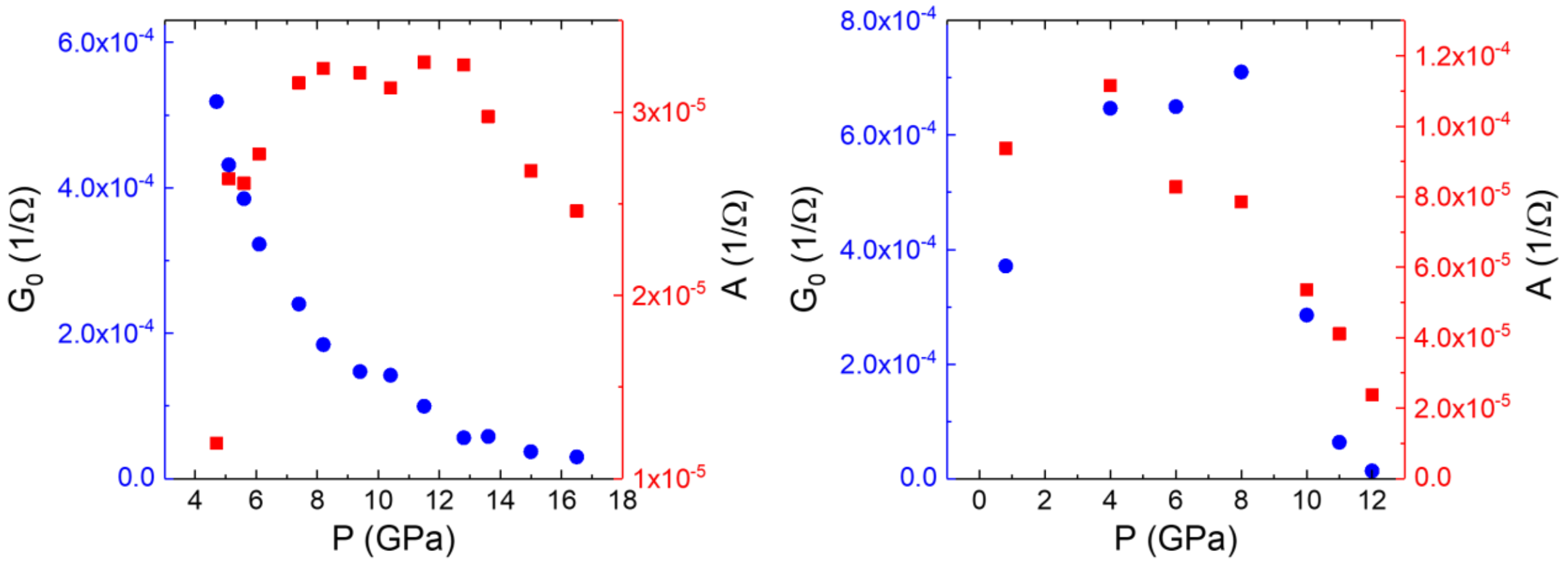

FIG. S3: $G_0$ and $A$ as a function of pressure extracted from fits to Eq. S1 for *Sample T* (left) and *Sample S* (right).

## III. ANALYSIS OF THE INSULATING REGIME

In the main text, we showed fitting for the insulating phases under ambient pressure to an activated insulator and under high pressure to an Anderson insulator. Fig.S4 shows the attempts to fit the curves to the opposite behavior. It can be seen that the fitting we showed in the main text is more satisfactory, supporting the idea of that the nature of the insulator changes under high pressure.

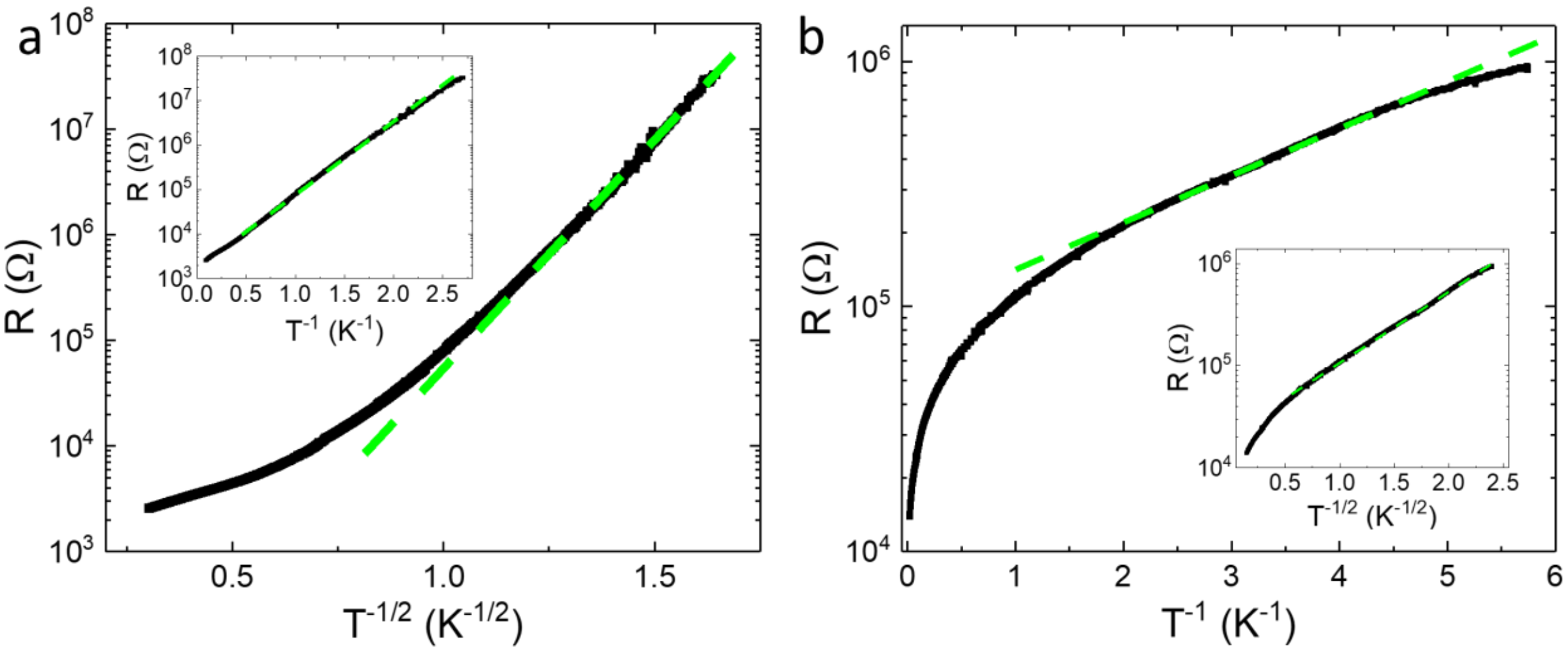

FIG. S4: **(a)** Resistance versus $T^{-1/2}$ for *Sample I* at ambient pressure. Inset: Resistance versus $1/T$ at ambient pressure. The comparison shows that simple activation, typical of a BI, more adequately describes the sample's behavior. **(b)** Resistance versus $1/T$ for *Sample S* at $P = 12$ GPa. Inset: Resistance versus $T^{-1/2}$ at $P = 12$ GPa. The comparison shows that Efros-Shklovskii VRH, typical of an Anderson insulator, more adequately describes the sample's behavior.

## IV. REVERSIBILITY

A question may arise as to the mechanical robustness of the film at high pressure. In particular, the transition to the fermionic insulator *I2* could be attributed to some defect in the sample, such as pressure-induced cracks. To rule out this possibility, we measured electronic transport as the sample was initially pressurized before being de-pressurized after reaching the *I2* phase. The results are depicted in = Fig.S5 where the dashed lines represent the measurements performed during depressurization. It is seen that the sample returns to an *M1* phase with signs of incipient superconductivity. Though the curves do not exactly trace back the behaviors before the pressure-de-pressure cycle, this shows that the process is quite reversible and that the phase transition to *I2* is a real electronic effect rather than due to permanent mechanical damage.

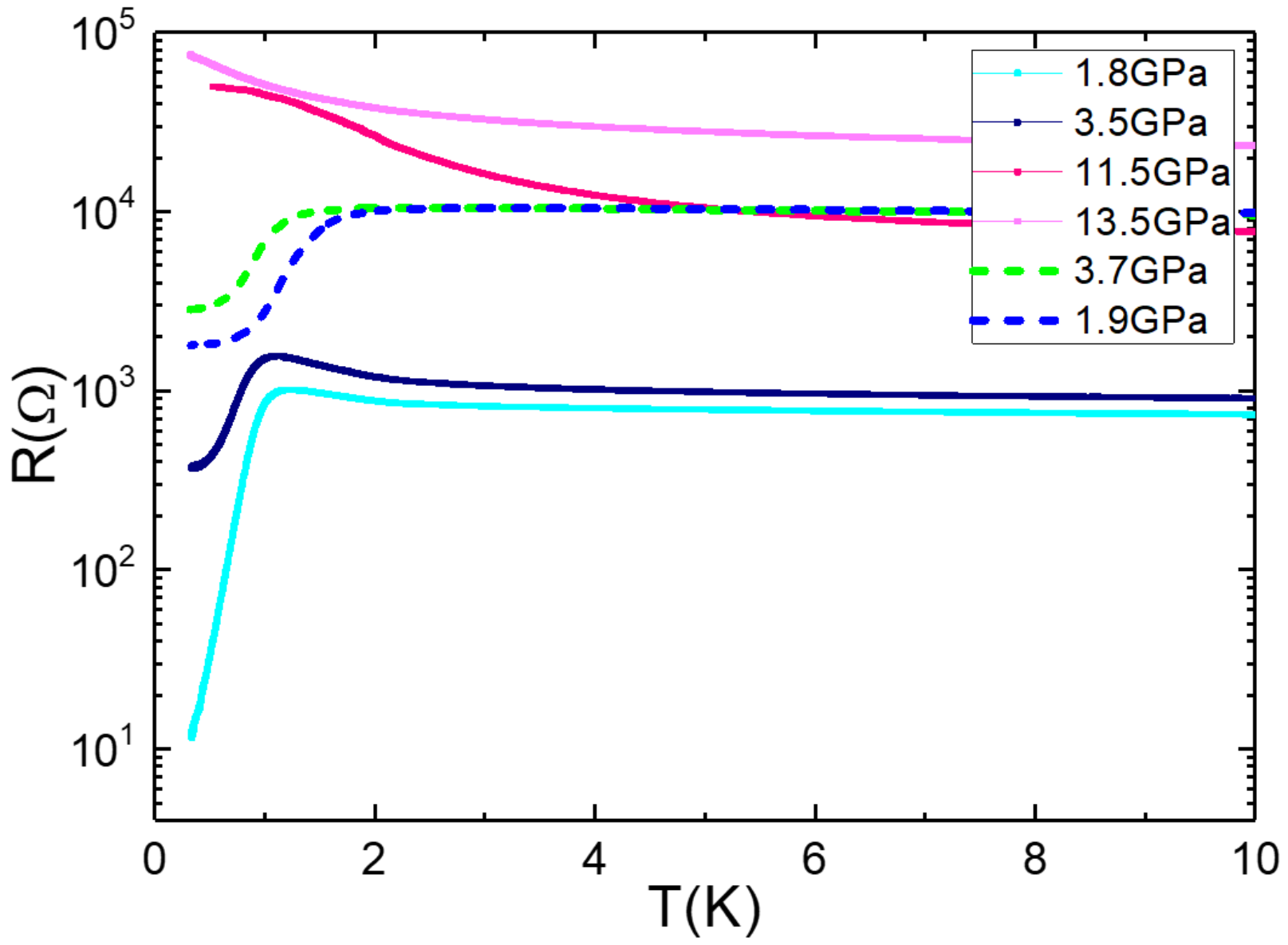

FIG. S5: Resistance versus temperature for *Sample I* which is of type *I1* at ambient pressure. Under pressure, it undergoes transitions to *S*, *M1*, *M2* and *I2* phases and has then been depressurized. The dashed lines were measured after decreasing the pressure.